\documentclass[a4paper,fleqn,usenatbib]{mnras}

\usepackage[T1]{fontenc}
\usepackage{ae,aecompl}

\usepackage{graphicx}	
\usepackage{amsmath}	
\usepackage{amssymb}	

\title[Luminosity function of ringed galaxies]{The luminosity function of ringed galaxies}

\author[D. V. Smirnov, V. P. Reshetnikov]{
	Daniil V. Smirnov, Vladimir P. Reshetnikov
	\thanks{E-mail: v.reshetnikov@spbu.ru}
	\\
	St.Petersburg State University, 7/9 Universitetskaya nab., St.Petersburg, 
	199034 Russia
}

\date{Accepted 2022. Received 2022; in original form 2022}

\pubyear{2022}

\begin{document}
	\label{firstpage}
	\pagerange{\pageref{firstpage}--\pageref{lastpage}}
	\maketitle
	
	\begin{abstract}
		
		We perform an analysis of the luminosity functions (LFs) of two types
		of ringed galaxies -- polar-ring galaxies and collisional ring
		galaxies -- using data from the Sloan Digital Sky Survey (SDSS). 
		Both classes of galaxies were formed as a result
		of interaction with their environment and they are very rare objects.
		We constructed LFs of galaxies by different methods and found their 
		approximations by the Schechter function. The luminosity functions of both 
		types of galaxies show a systematic fall-off at low luminosities.
		The polar structures around bright ($M_r \leq -20^m$) and red ($g-r > +0.8$)
		galaxies are about twice as common as around blue ones.
		The LF of collisional rings is 
		shifted towards brighter luminosities compared to polar-ring galaxies.
		We analysed the published data on the ringed galaxies in several deep 
		fields and confirmed the increase in their volume density with redshift:
		up to z$\sim$1 their density grows as $(1+z)^m$, where $m \gtrsim 5$.
		
	\end{abstract}
	
	
	
	
	\section{Introduction}
	
	The luminosity function (LF) of galaxies plays an essential
	role for extragalactic astronomy and it is one of the basic descriptions 
	of the galaxy population (e.g. \citealt{felten1977}; \citealt{bst1988}).
	LF is important to estimate the luminosity
	and baryonic densities of the Universe and to test models of
	galaxy formation and evolution (e.g. \citealt{fhp1998}; \citealt{blanton2003};
	\citealt{fk2022}).
	
	Knowledge of the luminosity function of any specific type of galaxies 
	makes it possible to estimate the 
	volume density of galaxies in a given luminosity interval and to estimate 
	the possible evolution of this density with redshift.
	In this work, we intend to study LF of two types
	of ringed galaxies -- polar-ring galaxies (PRGs) and collisional ring
	galaxies (CRGs). 
	
	PRGs are ``symbiotic'' objects consisting of two morphologically
	and kinematically decoupled systems. In such objects, the 
	central galaxy (of early-type, typically) is surrounded along the minor axis 
	by an extended ring or disc, often resembling a spiral galaxy (see example 
	in \autoref{fig:separation}.)
	The main formation scenarios of such galaxies are associated with external 
	influences: a major and minor merging, tidal accretion, 
	a cold accretion from cosmic filaments (see discussions in
	\citealt{iodice2015b}; \citealt{em2019}).
	
	The space density of PRGs is poorly constrained. Simple estimates show 
	that polar structures are observed in a few percent of S0 galaxies in 
	the nearby Universe (\citealt{swr1983}; \citealt{whit1990}).
	Close results were also obtained in \cite{rfo2011} (hereafter R11), in 
	which an attempt was made to estimate the luminosity function of PRGs.
	Comparison of the LFs for PRGs with different
	types of central galaxies shows that the polar structures around E/S0 
	galaxies are more common by about 3 times than those around spiral ones (R11).
	
	CPGs are the result of a recent collision between two galaxies, in which 
	a small galaxy is passing through the central region of a disc galaxy 
	(e.g. \citealt{as1996}; \citealt{struck2010}; \citealt{fernan2021}; 
	see example in \autoref{fig:separation}). 
	The prototypical low-redshift example of such objects is
	the famous ``Cartwheel'' galaxy. Like PRGs, CPGs are rare objects and 
	their volume density is not well known. Current estimates vary by several 
	times (e.g. \citealt{th1977}; \citealt{fm1986}).
	
	Thus, both types of galaxies are manifestations of relatively recent 
	processes of interaction and external accretion. According to numerical 
	simulations, the polar rings/discs can be stable for $\sim 10^9$ years
	(\citealt{bc2003}; \citealt{brook2008}), collisional ring phase is more short-lived 
	($\leq 0.5 \times 10^9$ years, \citealt{map2008}).
	Like other relics of interactions, ringed galaxies should show an 
	increase in volume density with increasing redshift (e.g. \citealt{abr1996}). 
	
	To study the evolution of the interaction/merger rate,
	different types of objects were used -- pairs of galaxies, mergers,
	galaxies with tidal structures, M\,51-type galaxies, etc.
	(e.g. \citealt{ls2009}; \citealt{bridge2010}; \citealt{rm2011};
	\citealt{lotz2011}; \citealt{duncan2019}; \citealt{pearson2022} and
	references therein). The results of these observational studies are 
	in general 
	agreement, but the use of new types of objects can help clarify the 
	dependence of the rate of interactions on time and on the 
	characteristics of galaxies. 
	
	Polar-ring and collisional ring galaxies
	are very expressive from a morphological point of view. They are easier to
	identify than other interaction relics (e.g. tidal bridges and tails, envelopes, etc.) 
	and, therefore, they may be good traces of the galaxy interaction rate.
	Previous statistics of ringed galaxies based on relatively small and 
	heterogeneous samples of objects seem to support the gradual increase in 
	the rate of interactions of galaxies with redshift
	(e.g. \citealt{lavery1996}; \citealt{resh1997}; \citealt{lavery2004}; 
	\citealt{reshdet2007}).
	Also, this conclusion is apparently supported by cosmological numerical 
	simulations (\citealt{edo2008}; \citealt{ela2018}).
	
	In recent years, the number of known nearby PRGs and CPGs has increased 
	significantly. This makes it actual to construct their LFs
	and to determine their local volume densities. 
	These data will allow to better understand the origin of these unique types 
	of galaxies, as well as to constrain possible evolution of their volume density.
	
	This paper is organised as follows. In the next section, we describe
	our samples of PRGs and CPGs. In Section~\ref{sec:LF}, we present
	our methods for constructing the LF of ringed galaxies and the results obtained. 
	Volume density evolution of galaxies is discussed in Section~\ref{sec:density}.
	
	Throughout this article, we adopt a standard flat $\Lambda$CDM
	cosmology with $\Omega_m$=0.3, $\Omega_{\Lambda}$=0.7, 
	$H_0$=70 km\,s$^{-1}$\,Mpc$^{-1}$. All magnitudes in the paper are given
	in the AB-system.
	
	\section{Samples of galaxies}
	
	The first step of our study is to construct a sample for both types of
	galaxies under analysis. There are several ways to approach this task: one can
	combine already established catalogues of objects under study or develop a new
	sample from the outset on the basis of some selection process. This selection
	can be done by a visual classification, by employing some numerical criteria 
	based on galaxy parameters or with the help of an image analysis technique.
	Due to the underlying complexity of galaxy classification visual inspection is
	still the superior approach although with toady's volumes of astronomical data it
	becomes	extremely time consuming to select galaxies purely by eye-looking.
	If the galaxies of interest are well restricted in terms of their light distribution
	or overall shape they can be selected based on their internal parameters, for instance
	Gini coefficient ($G$) and $M_{20}$ index are known to be a good indicator of galaxy's
	morphology \citep{lotz2004} while axial ratio is a good tool to select edge-on galaxies
	\citep{mitronova2004, EGIS}. With the ever increasing computational capacity of modern
	processors it is now possible to employ computer vision techniques with explicit algorithms
	\citep{ts2017} or machine learning methods \citep{yi2022, marchuk2022, EGIPS, vavilova2021,
	cheng2020}. Although ML is a very powerful tool that proved itself in a plethora of scenarios
	this technique has its own limitations, namely, a requirement of a relatively large
	representative training sample.
	
	It this study, we decided to build our samples by compiling different sources
	which looked for polar-ring and collisional ring galaxies or at least recorded them.
	Such approach allows us to take advantage of previous success in finding PRGs and
	CRGs across large regions of the sky. On the other hand, building a new sample on the basis
	of visual classification requires a lot of work much of which has already been done 
	(see references in \autoref{sec:prg_sample} and \autoref{sec:crg_sample}). It is also
	difficult to apply numerical selection criteria in case of these galaxies as PRGs and
	CRGs are poorly constrained by their structural parameters. Using machine learning is
	problematic too as the overall number of known polar and collisional rings is too
	small for an adequate training sample (see \autoref{tab:samples}). Another obstacle
	to the use of ML is a high diversity of apparent morphology for both types of galaxies,
	for instance SPRC-7 and SPRC-69 while both being the best candidates for PRGs, look very
	dissimilar due to a difference in viewing orientations and system configuration. This makes
	it almost impossible at the moment to construct a robust automatic algorithm for finding
	polar-ring and collisional ring galaxies.

	\subsection{Polar-ring galaxies}\label{sec:prg_sample}
	
	The sample of polar-ring galaxies under analysis in this work was derived 
	from several papers which catalogued PRGs across SDSS field of view. 
	
	The bulk of the sample consists of objects from the {\it SDSS-based
		Polar Ring Catalogue} by {\citet{moiseev2011}} (=\,SPRC).
	To construct this catalogue the authors utilised data from the Galaxy
	Zoo project \citep{gz1}. Using preliminary sample they managed to formulate a
	broad selection criteria for galaxies similar to already known PRGs in terms of
	Galaxy Zoo types. Authors looked through more than 40\,000 images of SDSS
	galaxies meeting this criteria and selected a total of 275 PRG candidates.
	The galaxies in the SPRC are divided into 4 groups: best candidates (70 objects),
	good candidates (115 objects), related objects (53 galaxies),
	possible face-on rings (37 galaxies).
	
	We included in the sample most of the ``best candidates'' from the
	SPRC. Also, based on their apparent morphology, we added a number 
	of ``good candidates'' (SPRC-71, 73, 77, 80, 84, 87, 90, 101, 132, 137, 
	142, 160, 161, 168).
	
	Another important source of PRGs is the paper by \citet{rm2019}
	who continued the hunt for polar-ring galaxies in the Galaxy Zoo
	data. Authors searched discussion boards for mentions of possible PRG candidates
	and carefully examined each case. After close inspection of SDSS images they were
	left with 31 new galaxies which are morphologically similar to the best 
	candidates from SPRC.
	
	Finally we added 5 well-known polar-ring galaxies
	from \citet{whit1990} ({\it Catalog of polar-ring galaxies}, PRC)
	which are covered by SDSS: PRC A-1, A-3, A-4, A-6, B-17. The PRC 
	is based on the search for PRGs on photographic plates. It
	lists 157 galaxies, from which only a small part are PRGs.
	
	\subsection{Collisional ring galaxies}\label{sec:crg_sample}
	
	To assemble our CRGs sample we have collected galaxies from several papers 
	which used different methods to select galaxies with collisional rings. 
	
	As a first step we used the well-known catalogue presented by
	\citet{madore2009}. In that paper authors revisited and reclassified all 
	ring galaxies from \textit{A catalogue of southern peculiar galaxies and 
		associations} \citep{am87} in order to select objects with crisp rings, 
	bearing footprints of recent or ongoing interaction. The second part of their
	catalogue comes from extensive literature search for previously studied ring 
	galaxies with similar morphology. Unfortunately, original catalogue by \cite{am87} 
	was not covered by the SDSS so only galaxies from the second
	part of Madore's list were included in our sample.
	
	Another group of CRGs in our sample comes from a detailed morphological catalogue
	of 14034 SDSS galaxies presented by \citet{nair2010}. 
	In that paper the authors performed a detailed visual 
	classification of nearby galaxies with $0.01<z<0.1$ and $g<16^m$.
	Besides numerical Hubble type for
	each galaxy the catalogue provides information on the presence of bars, lenses, 
	tails,
	rings and other features. In their paper authors outlined a group of ringed galaxies probably
	produced by a ``bulls-eye'' collision. They even found a presumably double collisional
	ring system SDSS J155308.66+540850.42. All 13 collisional galaxies from this
	source we added to our sample.
	
	Another approach was taken by \citet{ts2017} who developed a computer analysis 
	method for detecting ring galaxy candidates and applied it to the first data 
	release of Pan-STARRS \citep{PanSTARRS}. The employed algorithm
		analysed binary maps with different thresholds of galaxy images. If at some
		threshold the algorithm spots an area fully separated from the image's edge
		this galaxy is considered a CRG candidate (see \citet{ts2017} Section 2.2).
	Although such procedure has obvious limitations and accuracy issues, under
	human supervision it allows one to analyse large number of galaxies which
	would be impossible to classify manually. Later \citet{shamir2020} applied
	the same algorithm to SDSS DR14 images. All galaxies from these two catalogues
	labelled as collisional ring galaxies were included into our sample.
	
	Citizen science projects like the Galaxy Zoo are also a well-established way 
	to classify huge samples of galaxies obtained from digital sky surveys. 
	\citet{buta2017} took this route and properly classified almost 3700 galaxies
	identified as having ring structures by volunteers during the Galaxy Zoo 2 
	project \citep{gz2}. That paper presents a large sample of
		ring galaxies with a detailed morphological classification in terms of
		the CVRHS (Comprehensive de Vaucouleurs revised Hubble-Sandage) system.
	In addition Buta outlined a group of 20 objects called 
	``cataclysmic or encounter-driven rings'' to account for their collisional
	origin. We added these galaxies to our sample except for those included in 
	the SPRC.
	
\begin{figure*}
		\centering
		\includegraphics[width=.45\textwidth]{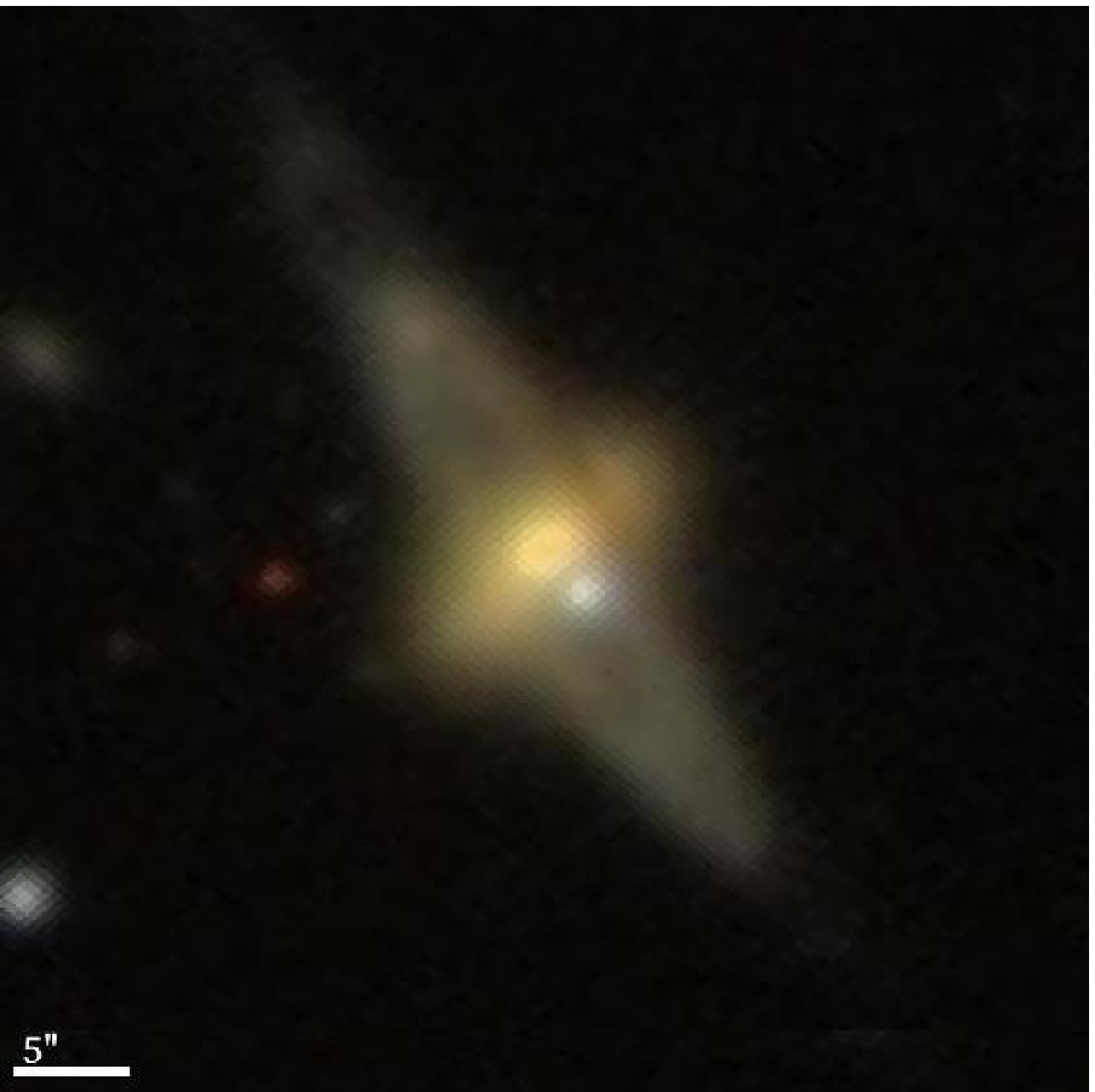}~~
		\includegraphics[width=.45\textwidth]{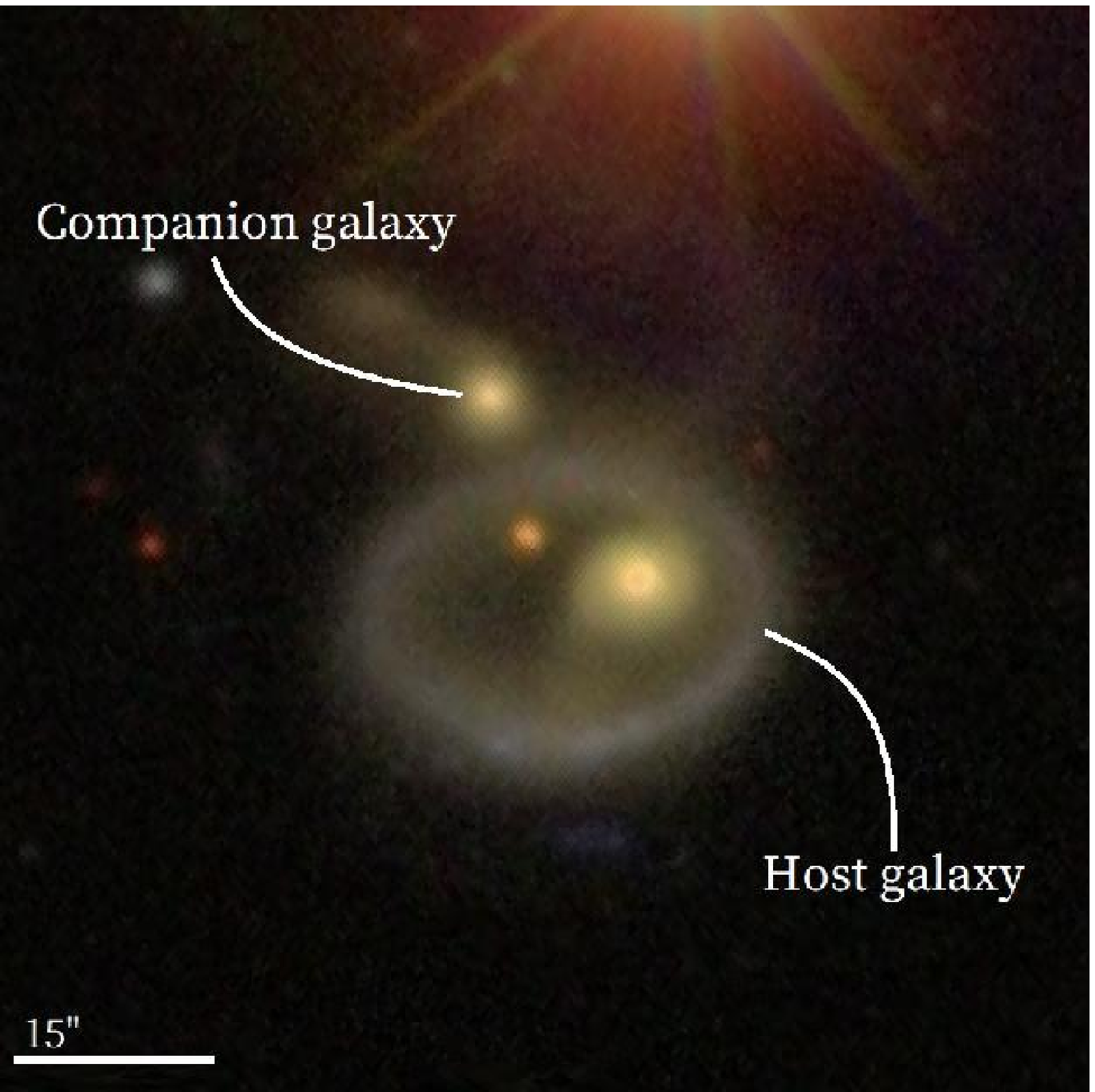}
		\caption{SDSS colour images of two objects under study. {\it Left}: 
			polar-ring galaxy SPRC-69 (\citealt{moiseev2011}). {\it Right}: 
			IIHz4 (No 365 in \citealt{buta2017}) showing the identification of a 
			ring host and a companion.}
		\label{fig:separation}
	\end{figure*}

	The next step was to distinguish ring hosts from possible companions/colliders
	as done 
	by \citet{madore2009}. For this we looked through SDSS colour images of galaxies 
	from \citet{ts2017}, \citet{shamir2020}, \cite{buta2017}, \citet{nair2010} and
	classified them as ``hosts'' or ``companions'' based on their morphology and 
	redshift if the latter was available. The sole aim of this procedure
	was to ensure that the derived data comes from a ring host as the angular separation
	between the interacting galaxies may be quite small. 
	An example of this procedure
	is presented in \autoref{fig:separation}.
	
	\subsection{Data}
	
	After the PRG and CRG samples were constructed, for each galaxy necessary data 
	from SDSS (DR16, \citealt{sdss16}) was extracted. This includes 
	dereddened \citep{schfin2011} apparent magnitudes in $r$ and $g$ bands as well 
	as spectroscopic redshifts. For galaxies whose redshift has not been measured 
	by SDSS we used values provided in the NED\footnote{NASA/IPAC Extragalactic 
		Database -- http://ned.ipac.caltech.edu} if present. In this study, we consider 
	only galaxies with spectroscopic redshifts as photometric ones may have large 
	uncertainties which can lead to incorrect estimation of LF. 
	
\begin{figure*}
		\centering
		\includegraphics[width=\textwidth]{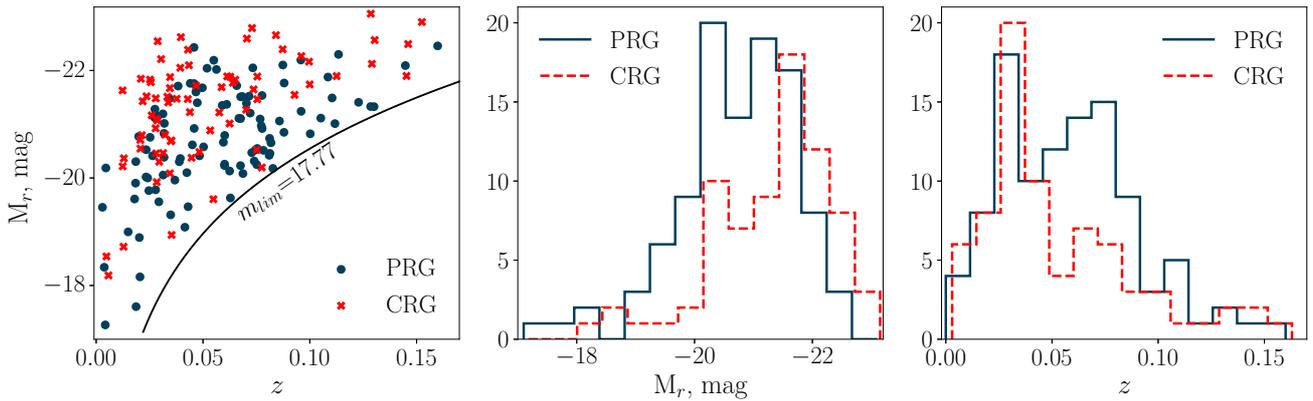}
		\caption{Left panel: Distribution of PRGs and CRGs in (M$_r$,z) 
			plane, black line shows limiting apparent magnitude. Middle and right panels 
			show the distribution of the samples galaxies by absolute magnitude and redshift
			respectively.}
		\label{fig:dist}
	\end{figure*}

	As the final selection
	criteria we applied $r$-band magnitude limit $m_{lim}=17.77$ following SDSS 
	Legacy Survey Target Selection algorithm. This restriction left 103 PRGs and 
	74 CRGs in consideration. \autoref{tab:samples} 
	summarises the final composition of our samples. The last column shows the number of objects left
	in consideration after all selection criteria were applied. Absolute magnitudes
	and colours of all galaxies	were corrected using the $k$-correction calculator by
	\citet{kcorr}.	Distributions of the main characteristics of the PRGs and CPGs in
	our samples are	presented in \autoref{fig:dist}.
	
	\begin{table}
		\caption{Description of the ringed galaxies samples used in our study}
		\label{tab:samples}
		\begin{tabular}{ccc}
			\hline
			Sample  &    Source       & Number of galaxies \\ \hline
			PRG & \citet{whit1990} & 5 \\
			& \citet{moiseev2011} & 75 \\
			& \citet{rm2019} & 23 \\
			& &$\Sigma\,$=\,103~~~~~~~\\ \\
			
			CRG     & \citet{madore2009} & 19 \\
			& \citet{nair2010} & 13 \\
			& \citet{buta2017} & 13 \\
			& \citet{ts2017} & 17 \\
			& \citet{shamir2020} & 12 \\
			& &$\Sigma\,$=\,74~~~~~~\\ 
			
			\hline
			
		\end{tabular}
	\end{table}
	
	\section{Luminosity function}\label{sec:LF}
	
	\subsection{Completeness of the samples}\label{sec:completeness}
	
	In order to investigate completeness of our samples we used classical 
	$\left<V/V_{max}\right>$ test originally developed by \citet{Schmidt1968}. 
	To implement this method one needs to calculate two values for each galaxy: 
	$V$ -- comoving volume of a sphere which radius is the distance to the object 
	and $V_{max}$ -- comoving volume of a sphere with radius corresponding to the 
	maximum distance the galaxy could have and still be included in our sample. 
	Assuming homogeneous distribution of objects in space, the expected value 
	of $\left<V/V_{max}\right>$ for a complete sample is 0.5. In our
	case, this value equals to 0.196$\pm$0.028 for polar-ring galaxies sample and 
	0.087 $\pm$0.034 for collisional ring galaxies sample. These results imply 
	that the samples are far from complete and the necessary corrections should 
	be applied.
	
	A number of techniques have been developed to make such corrections, we have 
	adopted method from \cite{hs1973} as the most consistent with the used 
	completeness test. Another advantage of this method is that it does not 
	require to decrease the number of galaxies in the sample or model the selection 
	process. To apply this correction one needs to tabulate
	$\left<V/V_{max}\right>$ over a suitable interval of $m_{lim}$ and then 
	determine what number of galaxies should be added to keep $\left<V/V_{max}\right>$ 
	close to 0.5. A total of 65 polar-ring galaxies are missing from our sample so 
	we need to multiply calculated space density and luminosity function by a 
	correction factor $\xi_{PRG} = 168/103 = 1.63$. Respective values for 
	collisional sample are 60 galaxies and $\xi_{CRG}=1.81$.
	
	We note that the main drawback of this method is its integral nature, i.e. only 
	the overall number of galaxies is corrected but not the relative number of 
	bright and faint galaxies. This means that only the normalisation of the LF 
	is affected but not its shape.
	
	\subsection{Methods description}\label{sec:methods}
	
	Through the years a lot of different techniques have been proposed to estimate 
	luminosity function of galaxies (see \citealt{johnston2011}). In this study, we
	adopted 3 classical  nonparametric methods: $1/V_{max}$ \citep{Schmidt1968},
	C$^{-}$ \citep{lb1971} and Cho\l{}oniewski method (\citealt{chol1986}). Although
	these methods are pretty old they have a number of
	important features: firstly, these techniques were repeatedly used and tested
	so their statistical properties are well studied (e.g. \citealt{willmer97};
	\citealt{tyi00}). The second advantage of using these methods is that they 
	were established to estimate LF and space density of small samples which is 
	the case in this study.	Of course, there are certain drawbacks associated with
	the use of these estimators: $1/V_{max}$ can overestimate faint end of luminosity
	function, original C$^-$ method does not provide normalisation and Cho\l{}oniewski's 
	method requires solving a nonlinear system of equations. However, there are ways
	to overcome such difficulties.
	
	In order to quantify the shape of derived LFs we have fitted them with an 
	analytical approximation. The most commonly used parametric form of LF was 
	proposed by \citet{Schechter}:
	\begin{align}\label{eq:Schechter_function}
		\phi(M) = 0.4\ln(10)\phi^*\left( 10^{0.4(M^*-M)} \right)^{\alpha+1}e^{-10^{0.4(M^*-M)}},
	\end{align}
	where $\phi^*$ is the normalisation, $M^*$ marks the position of the turn-off 
	point and $\alpha$ gives the logarithmic slope of LF at its faint end.
	
	\subsubsection{$1/
		V_{max}$ method}
	
	One of the reasons this method was adopted in our study is that it allows us 
	to compare results with previous attempts to estimate the luminosity function 
	taken by R11 without having to account for differences in methods.
	
	Implementation of this method goes as follows: firstly, for each galaxy in 
	sample we calculate value of $V_{max}$ as
	\begin{equation}
		V_{max}^i = \frac{\Omega}{4\pi}\int_{0}^{z_{max}(M_i)}\frac{dV}{dz}\,dz,
	\end{equation}
	where $z_{max}(M_i)$ is the redshift of $i$th galaxy at which its $r$-band 
	apparent magnitude equals $m_{lim}$ and $\Omega$ is the solid angle of the 
	sample. Then galaxies are binned by their absolute magnitude and value of
	differential LF in each bin can be written as
	\begin{equation}
		\phi(M) = \frac{\xi}{\Delta M}\sum_i \frac{1}{V^i_{max}},
	\end{equation}
	where the summation is over galaxies with $M_i\in[M-0.5\Delta M, M+0.5\Delta M]$.
	For both our samples $\Delta M=0{\fm}58$, $\Omega_{PRG}=11663$ $\Box^{\rm o}$
	(SDSS DR7 imaging area), $\Omega_{CRG}=14555$ $\Box^{\rm o}$ (SDSS DR14 imaging 
	area) and the values of $\xi$ are given above.
	
	\subsubsection{C$^-$ and Cho\l{}oniewski methods}
	
	The C$^-$ method of \cite{lb1971} uses the distribution of galaxies in 
	the $(M,z)$ plane to probe the luminosity function. For a sorted sample 
	under consideration one can calculate a C$^-$ value for $j$th galaxy as 
	the number of galaxies with $M<M_j$ and $z<z_{max}(M_j)$. Using these 
	numbers cumulative LF $\Phi(M)$ can be calculated via recurrence relation. 
	In this work we adopted modification of this method developed by 
	\cite{chol1987} as it simultaneously provides shape
	and normalisation of LF. 
	
	\citet{chol1986} method was applied in its original
	form.
	
	\subsection{Volume density estimators}\label{sec:density_methods}
	
	Three different approaches were taken to evaluate the mean volume density 
	of galaxies in our samples. 
	
	The first one is naive:
	\begin{equation}
		\bar{n}=\xi\frac{N}{V},
	\end{equation}
	where $N$ is the size of the sample and $V$ is its total volume. 
	
	The second is LF based and can be written as
	\begin{equation}
		n_{LF} = \int_{M_{1}}^{M_{2}} \phi(M)\,dM,
	\end{equation}
	where $M_1$ and $M_2$ give the luminosity interval in which the
	volume density is calculated.
	
	The last one is so called "EEP estimator" developed by \cite{eep1988}:
	\begin{equation}
		n_{EEP} = \frac{\xi}{V}\sum_0^N\frac{1}{s(z_i)}.
	\end{equation}
	Here $s(z)$ is a selection function defined as
	\begin{equation}
		s(z) = \frac{\int_{M_{1}}^{\min(M_{max}(z), M_2)}\phi(M)\,dM}{\int_{M_{1}}^{M_2}\phi(M)\,dM}
	\end{equation}
	where $M_{max}(z)$ is the faintest absolute magnitude detectable at redshift $z$.
	
	\subsection{Results}
	
\begin{figure*}
		\centering
		\includegraphics[width=\textwidth]{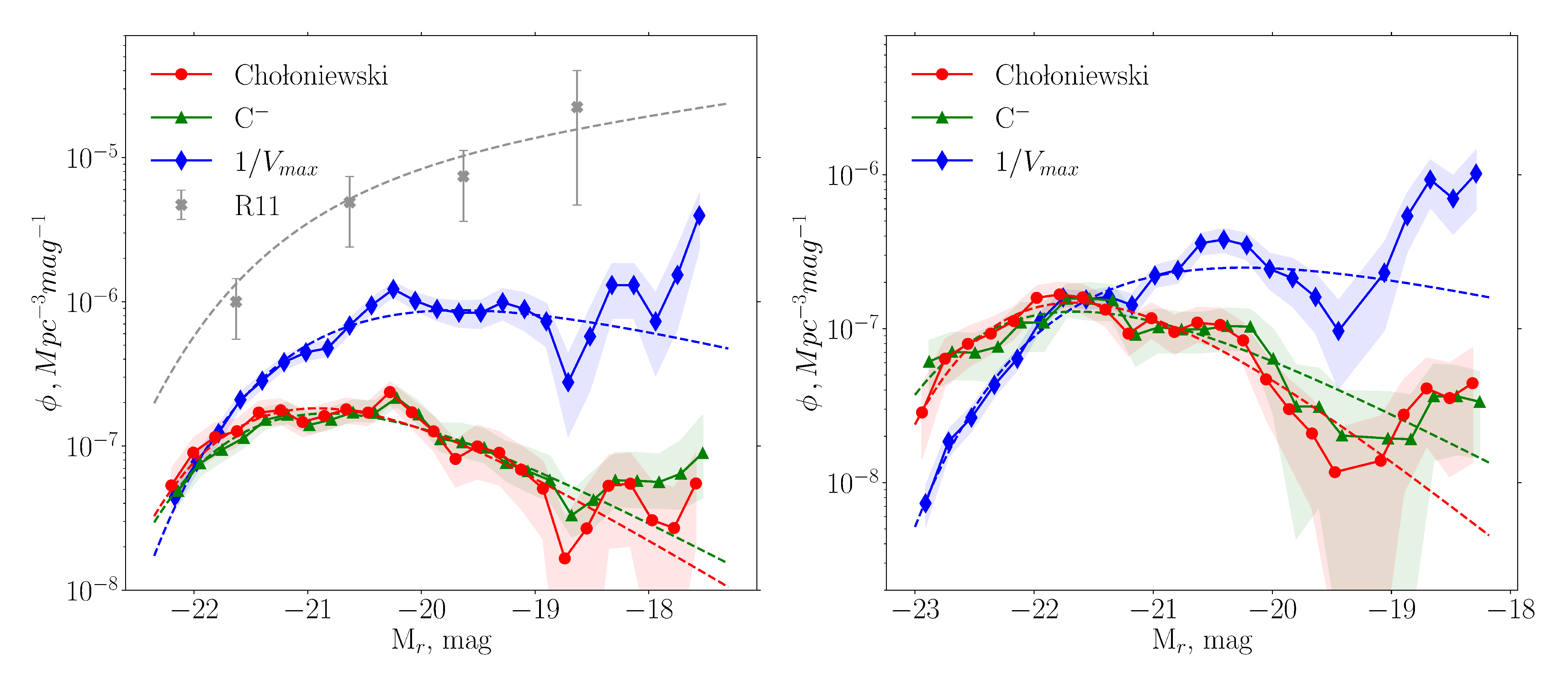}
		\caption{Luminosity function of PRGs \textit{(left panel)} and 
			CRGs \textit{(right panel)}. The diamonds, triangles, and circles represent 
			the results obtained with the $1/V_{max}$, C$^-$ and Cho\l{}oniewski methods,
			respectively. The filled areas show the 1$\sigma$ confidence regions of the LFs. 
			The dashed lines describe the analytic approximations with parameters listed 
			in \autoref{tab:fit_params}. The result from R11 is also added on the left panel.}
		\label{fig:LF_results}
	\end{figure*}

	\autoref{fig:LF_results} shows our results for the LF of polar-ring 
	and collisional ring galaxies. 
	Different symbols represent the results obtained by different methods described
	in Section~\ref{sec:methods}. The best Schechter fits are plotted with dashed 
	lines. Previous result for PRGs derived in R11 is also added to the left panel
	after transforming from the $B$ to $r$ filter according to \citet{cook2014}.
	Filled areas represent $1\sigma$ uncertainties, for C$^-$ and Cho\l{}oniewski methods
	they are calculated using bootstrap sampling, for $1/V_{max}$ we adopted an analytic
	formula by \citet{cond1989}. In order to better understand the shape of the LF we
	made 2 additional runs each time shifting luminosity bins by 0.19 mag. All three runs
	(one with original bins and two with shifted bins) are presented on \autoref{fig:LF_results}.
	
	It is apparent from this figure that the estimates produced by C$^-$ and 
	Cho\l{}oniewski methods show good agreement for both samples whereas results 
	from $1/V_{max}$ significantly deviates from them. 
	According to \citet{tyi00} such difference may indicate that a
	density inhomogeneity is present in our samples, as $1/V_{max}$ method is 
	affected by density fluctuations unlike the other two. 
	The best-fit parameters of the Schechter 
	function for each method are presented in \autoref{tab:fit_params}.
	
	As can be seen in \autoref{fig:LF_results}, the LF of PRGs from R11
	is much higher than the luminosity function found in this work.
	The main reasons for this may be both the incompleteness of the R11 sample 
	and its contamination with objects that are not  PRGs (e.g., mergers, interacting
	galaxies). The R11 sample is based on the PRC and it contains 
	nearby galaxies with $z \leq 0.05$ mostly (see fig.~7 in the SPRC).
	In a modest sample limited by a small spatial volume, the luminosity function 
	can be strongly distorted by the influence of spatial inhomogeneities.
	
	To check the influence of the redshift constraints on the LF found
	by the $1/V_{max}$ method,
	we built the LF of PRGs for subsamples with $z \leq 0.05$ 
	and $z < 0.03$ on the basis of our data. We found that as the redshift of 
	the sample decreases, the weak wing of the luminosity function 
	($M_r \geq -20^m$) rises significantly (by several times). 
	Therefore, the sensitivity of the $1/V_{max}$ method to the presence of spatial 
	fluctuations, as well as the incompleteness of the sample of galaxies, 
	apparently influenced the results of R11.
	
	\begin{table*}
		\caption{Schechter function best-fit parameters}
		\label{tab:fit_params}
		\begin{tabular}{ccccc}
			\hline
			Sample &    Method       & $\log_{10}(\phi^*,\,$Mpc$^{-3})$ & $M^*$           & $\alpha$ \\ \hline
			PRG & Cho\l{}oniewski & -6.28$\pm$0.02          			& -20.71$\pm$0.13 &  0.21$\pm$0.13 \\
			& C$^-$           & -6.31$\pm$0.02          			& -20.78$\pm$0.12 &  0.05$\pm$0.10 \\
			& $1/V_{max}$     & -5.66$_{-0.05}^{+0.04}$ 			& -20.49$\pm$0.11 & -0.53$\pm$0.13 \\ \\
			
			CPG & Cho\l{}oniewski & -6.42$\pm$0.03          			& -21.20$\pm$0.13 &  0.54$\pm$0.17 \\
			& C$^-$           & -6.42$\pm$0.03          			& -21.63$\pm$0.14 & 0.11$\pm$0.13 \\
			& $1/V_{max}$     & -6.23$_{-0.06}^{+0.05}$ 			& -21.17$\pm$0.14 & -0.58$\pm$0.18 \\
			
			\hline
			
		\end{tabular}
	\end{table*}
	
\begin{figure}
		\centering
		\includegraphics[width=0.5\textwidth]{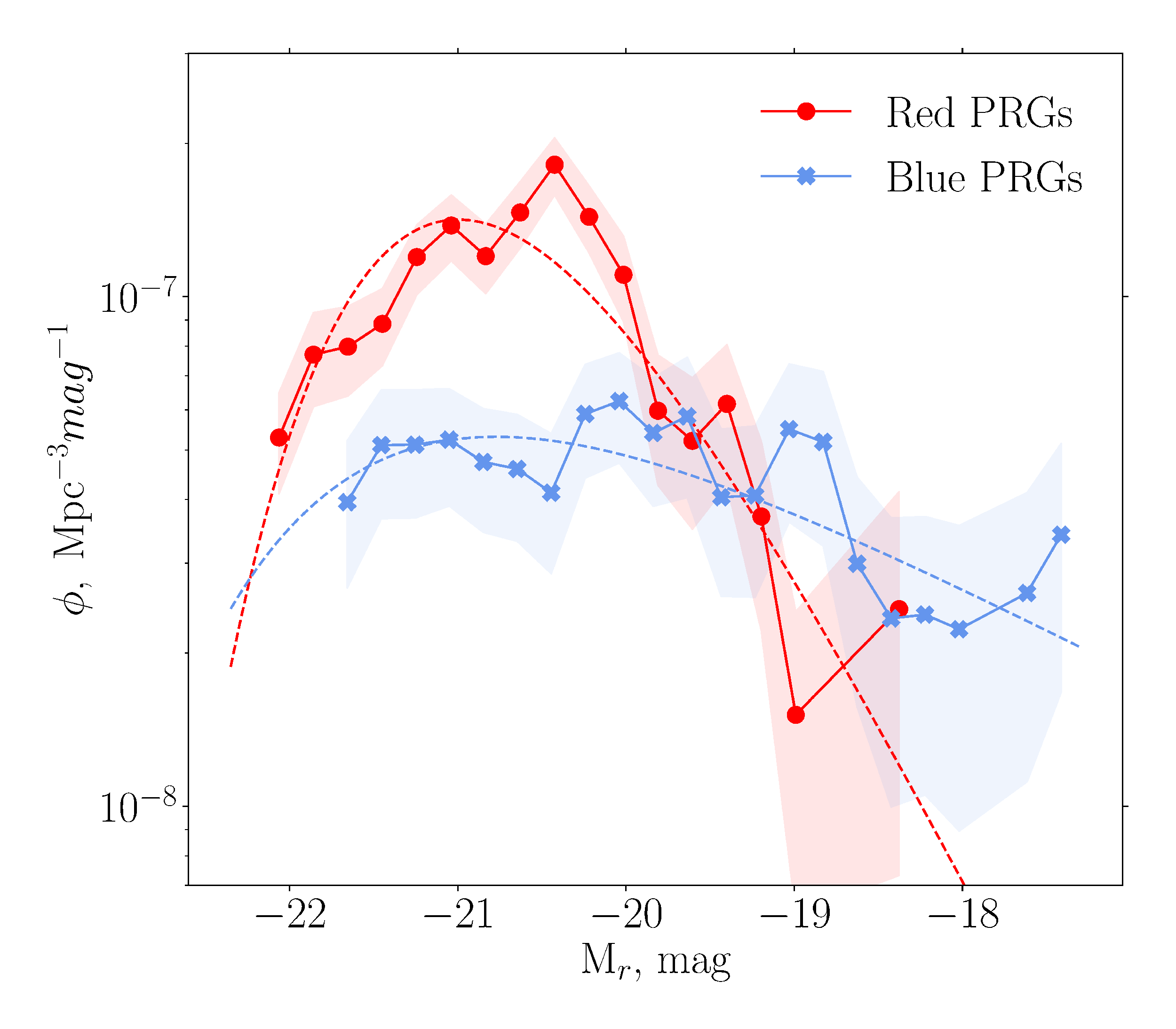}
		\caption{Mean luminosity functions of red PRGs (red circles) and blue PRGs 
			(light blue crosses). The filled areas show the
			1$\sigma$ confidence regions. Dashed lines represent best-fit Schechter functions.}
		\label{fig:LF_colour}
	\end{figure}

	\autoref{fig:LF_colour} shows LFs for red ($g-r > +0.8$, 65 galaxies) 
	and blue ($g-r < +0.8$, 38 galaxies) PRGs. The whole procedure described 
	above was applied to these subsamples, then the weighted mean of results 
	produced by inhomogeneity-insensitive methods (C$^-$ and Cho\l{}oniewski) 
	was calculated for both groups. As can be seen in the figure, the LFs of 
	the two types of galaxies are different. Red galaxies show a maximum and 
	decrease towards bright and faint luminosities, the LF of blue galaxies 
	looks more flat. Among bright objects, red galaxies dominate:
	in the luminosity range $-20^m \geq M_r \geq -22^m$, red galaxies are 
	found 1.7 times more frequent than blue ones. Among faint objects, the 
	contributions of blue and red PRGs are comparable.
	
\begin{figure}
		\centering
		\includegraphics[width=0.5\textwidth]{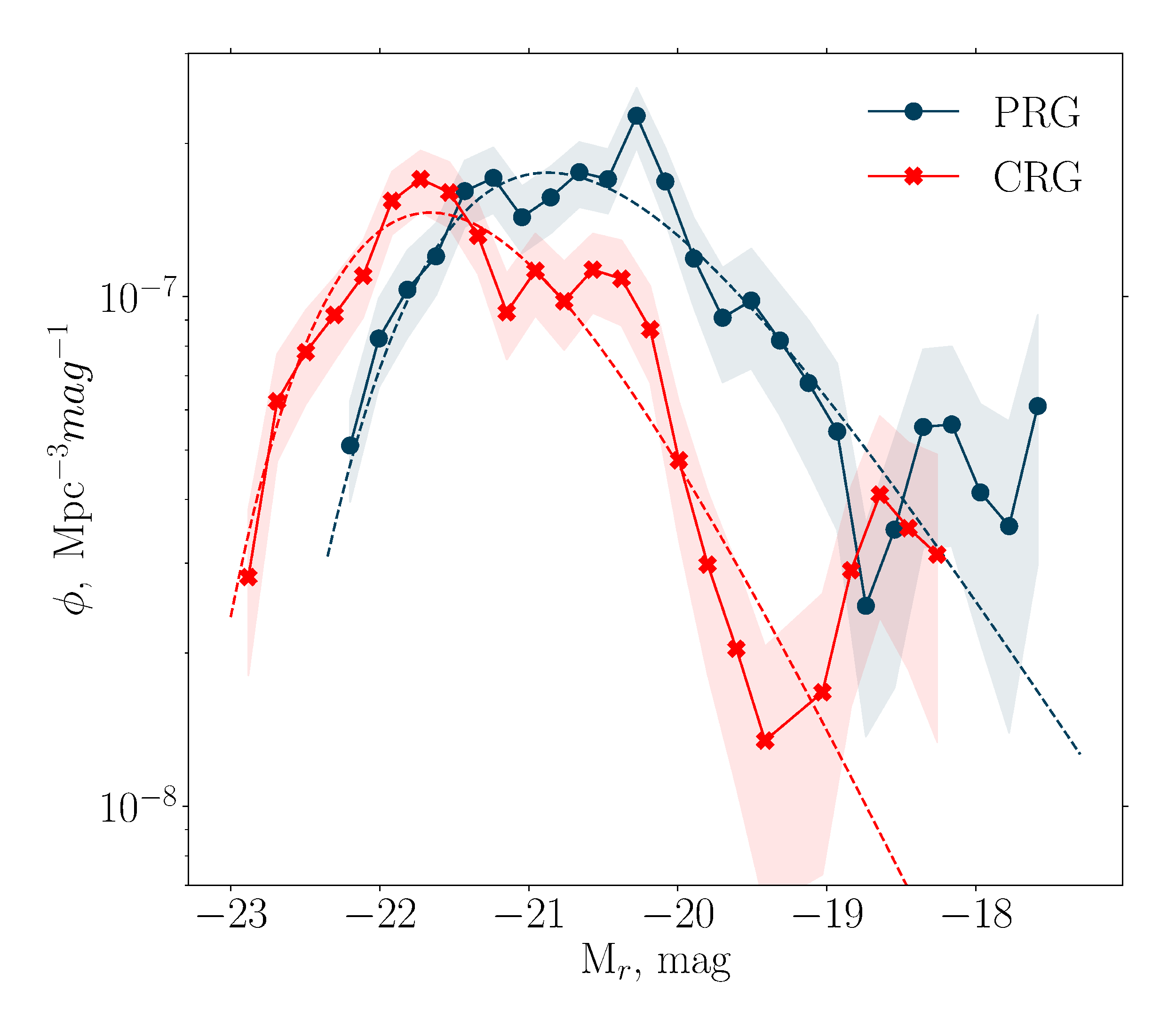}
		\caption{Mean luminosity functions of PRGs (blue circles) and CRGs (red crosses). 
			The filled areas show the 1$\sigma$ confidence regions. Dashed lines represent 
			best-fit Schechter functions.}
		\label{fig:LF_comparison}
	\end{figure}

	The mean LFs of PRGs and CPGs are compared in \autoref{fig:LF_comparison}.
	(As described above, the mean LFs were obtained as the weighted average of
	C$^-$ and Cho\l{}oniewski methods.) Both LFs exhibit similar behaviour with
	a gradual fall-off at low and high luminosities. Luminosity function of CPGs
	appears to be shifted to higher luminosities compared to the LF of 
	PRGs. This shift is also clearly visible from the data in \autoref{tab:fit_params} --
	characteristic absolute magnitudes $M^*$ for CPGs are brighter 
	by $0{\fm}5 - 0{\fm}8$ compared to PRGs.

	Once the luminosity functions were computed, volume density could be evaluated 
	employing estimators described in Section~\ref{sec:density_methods}. The 
	obtained values are presented in \autoref{tab:density_res}. We see that
	all estimates are consistent with each other. (In \autoref{tab:density_res}
	and below the results of the $1/V_{max}$ method are not used -- see discussion
	above). 
	
	\begin{table*}
		\caption{Volume density of PRGs and CRGs obtained by different estimators}
		\label{tab:density_res}
		\begin{tabular}{ccccc}
			\hline
			Sample &    LF Method    & $n_{EEP},10^{-7}\,$Mpc$^{-3}$ & $n_{LF},10^{-7}\,$Mpc$^{-3}$ & $\bar{n},10^{-7}\,$Mpc$^{-3}$ \\ \hline
			PRG & Cho\l{}oniewski & $6.74_{-0.60}^{+0.93}$      & $5.17\pm1.59$              &          \\
			& C$^-$           & $7.02_{-0.60}^{+0.86}$      & $5.23\pm1.78$              &  	$4.94$ \\ \\
			
			CPG & Cho\l{}oniewski & $4.04_{-0.16}^{+0.27}$      & $3.61\pm1.12$              &           \\
			& C$^-$           & $4.29_{-0.27}^{+0.45}$      & $3.57\pm1.24$              &  	$3.62$	\\
			
			\hline
		\end{tabular}
	\end{table*}
	
	The values we found -- $n_0(PRGs) \approx (5 - 7) \times 10^{-7}$\,Mpc$^{-3}$ and
	$n_0(CPGs) \approx 4 \times 10^{-7}$\,Mpc$^{-3}$ -- are noticeably smaller than
	those previously published.  According to our estimates, the relative fraction 
	of apparent PRGs is $\sim 10^{-4}$ of all galaxies; collisional rings are about 1.5
	times as rare as PRGs. 
	
	It should be noted that the value of $n_0$ for PRGs is the lower limit of 
	the real frequency of such galaxies. As can be clearly seen from the 
	PRC and SPRC catalogues, PRGs are predominantly identified in cases where the 
	polar structure is seen at a large angle to the line of sight, almost edge-on.
	As \citet{whit1990} noted, for every obvious PRG there are two other
	PRGs that are missed. Thus, the actual fraction of PRGs must be larger by
	approximately a factor of 3. Also, collision rings are easier to detect when they
	are	seen almost face-on. Therefore, the true volume density of CRGs should be slightly 
	higher than we have estimated.
	
	\section{Density evolution}\label{sec:density}
	
	If we know the local volume density of galaxies, we can estimate the redshift 
	evolution of this density. Implying
	some evolutionary model, for a given solid angle of the sky field $\Omega$
	and redshift range (from $z_1$ to $z_2$) one can calculate the expected number 
	of objects as
	\begin{equation}
		N_{exp} = \Omega\int_{z_1}^{z_2}n(z)\,\frac{dV}{dz}dz,
	\end{equation}
	where $dV$ is a comoving volume element.
	
	Volume density evolution is usually described by a simple 
	power law $n(z) = n_0 \cdot (1+z)^m$, where $n_0$ is
	the local density of objects under consideration. Then, comparing 
	$N_{exp}$ to the observed number of objects in the field,
	the $m$ exponent can be evaluated.
	
	\subsection{Polar-ring galaxies}\label{sec:density_PRGs}
	
	Up to date only three distant polar-ring galaxy candidates are known located 
	in the Hubble Deep Field North (HDF-N) and the Hubble Ultra-Deep Field (HUDF); 
	none were identified in the Hubble Deep Field South (\citealt{resh1997}; 
	\citealt{reshdet2007}). 
	
	HDF-N contains two candidates to PRGs. One galaxy (HDF-N 2-809) in its 
	morphology is similar to the nearby polar-ring galaxy NGC\,4650A, 
	and the second (HDF-N 2-906) resembles NGC\,2685 (\citealt{resh1997}). 
	The first galaxy is at redshift $z = 0.71$, the second one is at redshift
	$z = 1.24$ (\citealt{yang2014}). The third galaxy is in the HUDF (HUDF 1619,
	see \citealt{reshdet2007}) at redshift $z = 1.30$ (\citealt{rafelski2015}).
	
	The total area of three deep fields (HDF-N + HDF-S + HUDF)
	is $1.89\times10^{-6}$\,sr. Adopting the local volume density of PRGs 
	$n_0(PRGs) = 6 \times 10^{-7}$\,Mpc$^{-3}$ (\autoref{tab:density_res}),
	we found that in the case of no evolution ($m=0$) the expected number
	of PRGs over the redshift interval of $0 \leq z \leq 1.3$
	in the directions of three deep fields is only 0.02.
	Assuming Poisson errors, observed number of polar rings is
	consistent with an exponent value of $m=7.0_{-1.2}^{+0.6}$ 
	This means a very steep increase of PRGs volume density with redshift.
	
	\citet{fgb2010} found a candidate for PRG in the Subaru Deep Field (SDF)
	at $z = 0.061$. Assuming $\Omega(SDF) = 7.62\times10^{-5}$\,sr, $z \leq 0.1$ and 
	$m = 0$, we determined the expected number of PRGs as 0.001.
	In order to find one PRG in the direction of SDF, it is necessary to accept 
	an extremely high rate of evolution with $m >> 10$.
	Presence of one candidate for PRG in the VST Deep Field (\citealt{iodice2015a})
	at $z = 0.051$ also leads to an estimate of $m >> 10$.
	
	Therefore, the current very scarce statistics on the frequency of PRGs provide 
	indications of the possibility of a rapid evolution of the spatial density
	of such galaxies up to $z \sim 1$.
	
	\subsection{Collisional ring galaxies}\label{sec:density_CRGs}
	
	Statistics of high-redshift collisional ring galaxies is much richer. 
	\cite{lavery1996} identified 7 P-type
	rings in the \textit{HST} images of the Tucana dwarf galaxy. Additional 
	25 CRGs were spotted in deep images
	from the \textit{HST} archives by \cite{lavery2004}. Later, \cite{elmegreen2006} 
	listed 24 collisional rings and 15 ``bent chains'' found in GOODS and GEMS 
	fields. 
	
	Combining these 3 sources and discarding bent chains (as
	they morphologically differ from our local CRG sample) we get 56 galaxies 
	with $0.0< z \le1.4$ located in the total area of $1.39\times10^{-4}$ sr. 
	Corresponding local density $n_0 = 4\times10^{-7}\,$Mpc$^{-3}$ (\autoref{tab:density_res}). In a
	nonevolving case the expected number of CRGs in these fields is 1 and 
	the value of $m$ corresponding to the observed number of objects is 
	$5.2_{-0.2}^{+0.2}$ which is consistent with results obtained 
	by \cite{lavery2004}.
	
	It should be noted that although local densities obtained by C$^{-}$ and 
	Cho\l{}oniewski's methods are somewhat different, these variations do not 
	affect $m$ significantly. If we pick any of four estimations, the value 
	of $m$ would vary by less than 5\%.
	
	\section{Conclusions}
	
	Based on the SDSS data, we constructed the luminosity functions for two 
	types of ringed galaxies -- polar-ring galaxies and collisional ring galaxies. 
	Both types of galaxies are natural consequences of galactic 
	collisions (CPGs), interactions and external accretion of matter (PRGs). 
	Consequently, the statistics of such objects at different $z$ can be used to 
	estimate the rate of galaxy interactions in different epochs.
	
	We used different approaches to evaluate LF. Two of them (C$^-$ and 
	Cho\l{}oniewski methods) showed good agreement, method $1/V_{max}$, 
	sensitive to density inhomogeneities, gave overestimated LF values.
	Our main conclusions are based on the results of the first two approaches.
	
	-- LFs of ringed galaxies have global maxima and falling wings at high and 
	low luminosities (\autoref{fig:LF_comparison}). 
	LF of CRGs is shifted towards higher luminosities compared 
	to PRGs. The decrease in the number of PRGs and CRGs at low luminosities 
	may be due to the fact that the formation of large-scale optical polar 
	structures and extended collisional rings is more likely in non-dwarf galaxies.
	As for PRGs, our conclusion is in general agreement with the results of \citet{mis}, 
	who found that the fraction of kinematically misaligned galaxies declines to both 
	low and high mass end.
	
	-- Polar structures are more common in red (early-type) galaxies 
	compared to blue ones (\autoref{fig:LF_colour}). This is consistent with 
	direct estimates of the morphological types of PRGs (e.g. \citealt{whit1991}).
	This result looks natural, since the polar structures around spiral galaxies 
	with extended gaseous discs should exist for a shorter time compared to gas-free 
	red ones. (However, polar structures also exist around spiral galaxies -- 
	see discussion in \citealt{moiseev2014}).
	
	-- Very poor statistics of distant PRGs does not contradict the assumption 
	of a rapid evolution of their volume density to $z \sim 1$ (Sect.~\ref{sec:density_PRGs}).
	Since the rate of interactions and mergers of galaxies increases with redshift
	(e.g. \citealt{con2009}),
	our result is in agreement with the standard assumption that formation of 
	PRGs is associated with their interactions with the environment: major and minor
	merging, tidal accretion of matter from nearby galaxy, infall of gas from intergalactic
	space (e.g. \citealt{bekki1997}; \citealt{rs1997}; \citealt{maccio2006}).
	
	-- Current statistics of CRGs confirms the rapid increase in their volume density 
	towards $z \sim 1$ (Sect.~\ref{sec:density_CRGs}). This rapid increase may 
	reflect an increase in the rate of high-speed collisions of galaxies, leading to the 
	formation of such objects. On the other hand, this growth is apparently not 
	monotonic. According to \citet{yuan2020}, the volume vensity of massive CPGs
	at $z \approx 2$  can be comparable to the same density in the nearby Universe.
	(This decline may be due to a decreased fraction of large spiral galaxies at high redshift.)
	Non-monotonic behavior of the interaction/merger rate evolution is consistent with the 
	data obtained for other types of objects (e.g.
	\citealt{con2008,con2022}); however, the statistics of CPGs at highest $z$ is still 
	insufficient for meaningful conclusions.
	
	The results of our work show that polar-ring and collisional ring
	galaxies can be useful indicators of the rate of interactions at high 
	redshifts. Nevertheless, for the successful use of these indicators, it 
	is necessary to significantly increase the statistics of such objects
	in the local Universe and at different $z$. 
	
	Modern wide-field sky surveys (SDSS, Pan-STARRS, Legacy, etc.) provide 
	morphological and photometric information on millions of galaxies,
	which requires computer analysis methods to select and recognize 
	the desired objects. Such studies are now being carried out 
	(for example, \citealt{ts2017}) and it is hoped that in the coming years 
	our knowledge of nearby galaxies (including PRGs and CPGs) will improve 
	significantly.
	
	The main tool for studying the distant Universe in the coming years 
	is the James Webb Space Telescope. The operation of this telescope 
	will make it possible to detect a large number of ringed galaxies at high 
	redshifts, similar to the one discovered by \citet{yuan2020} at $z=2.19$.
	The combination of these data will make it possible to trace the 
	evolution of ringed galaxies over cosmological times.

	\section*{Acknowledgements}
	
	The authors thank an anonymous referee for constructive comments and 
	suggestions.
	
	Funding for the Sloan Digital Sky Survey IV has been provided by the 
	Alfred P. Sloan Foundation, the U.S. Department of Energy Office of 
	Science, and the Participating Institutions. 
	
	SDSS-IV acknowledges support and resources from the Center for High 
	Performance Computing  at the University of Utah. The SDSS 
	website is www.sdss.org.
	
	SDSS-IV is managed by the Astrophysical Research Consortium 
	for the Participating Institutions of the SDSS Collaboration including 
	the Brazilian Participation Group, the Carnegie Institution for Science, 
	Carnegie Mellon University, Center for Astrophysics | Harvard \& 
	Smithsonian, the Chilean Participation Group, the French Participation Group, 
	Instituto de Astrof\'isica de Canarias, The Johns Hopkins 
	University, Kavli Institute for the Physics and Mathematics of the 
	Universe (IPMU) / University of Tokyo, the Korean Participation Group, 
	Lawrence Berkeley National Laboratory, Leibniz Institut f\"ur Astrophysik 
	Potsdam (AIP),  Max-Planck-Institut f\"ur Astronomie (MPIA Heidelberg), 
	Max-Planck-Institut f\"ur Astrophysik (MPA Garching), 
	Max-Planck-Institut f\"ur Extraterrestrische Physik (MPE), 
	National Astronomical Observatories of China, New Mexico State University, 
	New York University, University of Notre Dame, Observat\'ario 
	Nacional / MCTI, The Ohio State University, Pennsylvania State 
	University, Shanghai Astronomical Observatory, United 
	Kingdom Participation Group, Universidad Nacional Aut\'onoma 
	de M\'exico, University of Arizona, University of Colorado Boulder, 
	University of Oxford, University of Portsmouth, University of Utah, 
	University of Virginia, University of Washington, University of 
	Wisconsin, Vanderbilt University, and Yale University.
	
	This research has made use of the NASA/IPAC Extragalactic Database, which 
	is funded by the National Aeronautics and Space Administration and 
	operated by the California Institute of Technology.

	\section*{Data availability}
	
	The data underlying this article will be shared on reasonable request to the 
	corresponding author.
	
	

	\bibliographystyle{mnras}
	\bibliography{art}
	
	\label{lastpage}
\end{document}